\documentclass[12pt,preprint]{aastex}

\newcommand{\teff}{T$_{\rm eff}$}
\newcommand{\logg}{log $g$}
\newcommand{\msun}{M$_{\odot}$}

\shorttitle{Draco 119}
\shortauthors{Fulbright et al.}

\begin{document}

\title{Draco 119: A Remarkable Heavy Element-deficient Giant\footnote{Based on data obtained at the W. M. Keck Observatory, which is operated as a scientific partnership among the California Institute of Technology, the University of California, and NASA, and was made possible by the generous financial support of the W. M. Keck Foundation.}}

\author{Jon. P. Fulbright}
\affil{Observatories of the Carnegie Institution of Washington, 813 Santa Barbara St., Pasadena, CA 91101}
\email{jfulb@ociw.edu}

\author{R. Michael Rich}
\affil{Division of Astronomy, Department of Physics and Astronomy, UCLA, Los Angeles, CA  90095-1562}
\email{rmr@astro.ucla.edu}

\author{Sandra Castro}
\affil{Infrared Processing and Analysis Center, Caltech, 770 S. Wilson Blvd., Pasadena, CA  91125 and The European Southern Observatory, Karl-Schwarzschild-Str. 2, Garching, D-85748}
\email{smc@astro.caltech.edu}

\begin{abstract}
We report the abundance analysis of new high S/N spectra of the most metal-poor
([Fe/H] $= -2.95$) star presently 
known to be a member of a dwarf galaxy, the Draco 
dSph red giant, D119.  No absorption lines for elements heavier than Ni are
detected in two Keck HIRES spectra covering the $\lambda\lambda$ 3850--6655 
\AA{} wavelength range, phenomenon not previously noted in 
any other metal-poor star.
We present upper limits for several heavy element 
abundances.  The most stringent limits, based on the non-detection of 
\ion{Sr}{2} and \ion{Ba}{2} lines, indicate that 
the total s- and r-process enrichment of D119 is at least 100 times smaller 
than Galactic stars of similar metallicity.  The light element abundances
are consistent with the star having formed out of material enciched primarily by
massive Type II supernovae (M $> 20$--$25$ M$_{\odot}$).  If this is the case, 
we are forced to 
conclude that massive, metal-poor Type II supernovae did not contribute to
the r-process in the proto-Draco environment.  
We compare the abundance pattern observed in D119 to current predictions of 
prompt
enrichement and pair-instability supernovae and find that the model predictions 
fail by an order or maginitude or more for many elements.
\end{abstract}


\section{Introduction}

The elements heavier than the iron group are believed to be produced     
primarily by neutron-capture processes.  In their landmark work on
the origin of the elements, \citep{bbfh} introduced the idea
that the production of heavy elements depends on whether the rate of neutron
capture is fast relative to the beta-decay timescale of
the resulting nucleus (``r-process'') or slow
relative to beta-decay (``s-process).

The exact astrophysical sites of these processes have been the basis of
much research.  \citet{sc03} and \citet{t02} review
the current understanding of the origin of the neutron-capture elements.
The s-process is generally accepted to occur in two sites:  
the He-burning cores of massive stars and the thermally pulsing He shells 
of AGB stars.  However, the specific site of the r-process is an unsolved
problem, but 
it is strongly suspected that Type II supernovae play an important role.  
\citet{t02} point out four possible sites, three of which occur
in Type II supernovae explosions.  

The mass range of Type II supernova progenitors that host the r-process
is also under debate.  \citet{m92}, \citet{t99} and \citet{w03}
propose that 8--10 \msun{} progenitors are the predominant contributors
to the r-process.  This mass range provides the best fit between their
r-process production models and the observed distribution of 
neutron-capture elements as a function of [Fe/H].   
Alternatively, \citet{t00} suggest that the largest
contributors to the r-process are 20--25 \msun{} progenitors. 
They do not place an upper mass limit to r-process contributors,
but they do not believe that the low mass Type II supernovae are significant
contributors to the r-process.  

Additionally, it has been suggested that an additional ``prompt'' enrichment 
episode is
necessary to enrich the protogalactic medium before the formation of 
metal-poor stars.  \citet{qw01}, \citet{qw02} and \citet{q03} put forth
a multi-component model for early chemical evolution.  The three components 
of the model are a ``prompt'' (P) component that represents the contribution 
from very high mass ($M > 100$ \msun) Population III stars, plus ``high''- 
and ``low''-frequency (H and L) enrichment events, both identified with Type II 
supernova.  The prediction for the prompt enrichment contains contributions
of the light neutron-capture elements Sr, Y and Zr.

A possible source of any prompt enrichment episode may have been a population
of zero-metallicity stars.
\citet{hw02} and \citet{un02} provide predictions of the yields of a 
first generation 
of very massive (100 to 300 M$_\odot$) stars.  These stars explode as pair
instability supernovae, and produce a distinctive nucleosynthesis pattern.
These supernovae are not predicted to produce large quantities of the
neutron-capture elements.

In this paper, we describe our analysis of the red giant star Draco 119 
\citep{s79}.  The star is the most metal poor in the Keck/HIRES sample
of \citet{sbs} which reports analysis of a relatively
low S/N ($\sim 25$) red spectrum of the giant.  Their reported [Fe/H]=$-2.97$ 
placed the star in an interesting range of abundance: nearly 1 dex more 
metal poor than any Galactic globular cluster, and as metal poor as the
point where the extreme Galactic halo shows striking new abundance
trends \citep{m95,rnb,j02}.   The star was targeted for observation
by one of us (RMR) to discover whether such a star would fit into the
extrapolation of the abundance trends for Galactic globular cluster giants,
or would rather show abundance patterns characteristic of the
ultra metal-poor halo stars. 

Even after detailed abundances have been analyzed for some 40 additional
giants in every Milky Way dwarf spheroidal galaxy except for Leo II,
D119 remains to date as the most metal-poor star presently documented
in any Milky Way dwarf spheroidal galaxy \citep{sbs, b00, scs, vlt03, mw03}.

Compared to the effort of \citet{sbs} we have invested
considerably more integration time and pushed blueward, yielding
two spectra of considerably higher S/N.  In the course
of analyzing this star, we discovered the extreme
deficiency of neutron-capture elements in this star, as immediately evidenced 
by the lack of a detectable Ba II 4554\AA\ line, the first ever 
such case in a metal-poor cool giant.

Given the unique nature of D119, we compare it to the bright,
well studied
halo star HD~88609 which shares similar stellar parameters,
and to BD~+80~245, a low-alpha metal-poor halo star \citep{c97, i03}
with low [Ba/Fe], yet still not as extreme as D119.  We finally compare the 
observed abundance distribution of D119 to predictions of early enrichment 
from a range of supernova models.

\section{Observations and Abundance Analysis}

The data are obtained using
two settings of the HIRES spectrograph \citep{vogt} on 
the Keck I telescope.  For both settings a slit returning a resolution of 
45000 was used.  The first setting, used on 6 July 2000, covers the wavelength 
range from $\lambda\lambda$ 3850--6280 \AA.  Five 3000 s exposures were taken 
with this setting.  The second setting, used on 29 June 2001, covers the 
wavelength range of $\lambda\lambda$ 4225--6655 \AA, with some gaps in the 
wavelength coverage.  Seven 3000 s exposures were taken with this setting.  
The data were reduced using the MAKEE\footnote{MAKEE was developed by T. A. 
Barlow specifically for reduction of Keck HIRES data. It is freely available on 
the world wide web at the Keck Observatory home page; 
http://www2.keck.hawaii.edu:3636/.} data reduction package.
The final signal-to-noise (S/N) varies as a function of wavelength, reaching 
over 60 per pixel in the red, but falling to $\sim 5$ per pixel in the blue.
The S/N at the $\lambda$ 4077 \AA{} and $\lambda$ 4215 \AA{} Sr II lines is 
about 7 per pixel, while the S/N at the $\lambda$4554 \AA{} and 
$\lambda$4934 \AA{} Ba II lines is about 35 per pixel.  

We checked the star for signs of binarity by comparing the observed
heliocentric radial velocities.  Matt Shetrone provided us with the 
28 August 1997 HIRES spectrum used in \citet{sbs}.  The mean 
radial velocity for D119 was measured to be $-293.9 \pm 0.1$ km/s.  The 
difference in radial velocity between the July 2000 and June 2001 spectrum is 
$-0.2 \pm 0.1$ km/s, while the difference in radial velocity between the 
August 1997 and June 2001 is $+0.2 \pm 0.3$ km/s.  We do not detect any large 
radial velocity variations over the nearly four years spanned by the spectra.

The initial line list was taken from \citet{j02},
which studied similar metallicity giants in the Galactic halo.  
Additional lines were added from 
\citet{jb02}, \citet{f00} and sources therein.
Hyperfine splitting was taken into account for 
\ion{Na}{1} D, \ion{Al}{1}, \ion{Mn}{1}, \ion{Co}{1}, 
\ion{Sr}{2}, \ion{Ba}{2}, and \ion{Eu}{2} lines.  
The C abundance of the two stars were determined by CH lines in the 
G-band using a line list provided by Andy McWilliam based on \citet{b87}.  
The line list and equivalent width measurements for both stars 
are given in Table 1. 

Many of the interesting absorption lines in D119 are very 
weak or undetectable in our spectra.  To place meaningful upper limits on 
the abundances of these elements, it is necessary to calculate 
upper limits for the equivalent widths of these lines.
The full Fe line list contains nearly 200 lines,
but only 96 were used in the final abundance analysis.  
We reject many of the remaining lines because they are too weak to measure.
The Fe abundance determined from the measured Fe lines can then be used
to predict the equivalent widths of the weaker, unmeasured, Fe lines.
By comparing the estimated equivalent widths of
these lines against the observed spectrum, we estimate the detection
limit as a function of wavelength and S/N.

The value of this limit is roughly twice that given by \citet{c88}
for the measurement error of the equivalent width of a line. 
This ``two-sigma limit'' allows for the estimation of the
measurement limit in regions where there are no Fe lines.
The two-sigma limit is $\sim 5$ m\AA{} in the red, but increases
to $\sim 20$--$30$ m\AA{} in the blue.

To compare D119 to similar stars in the Galactic halo, we have re-analyzed 
the well-studied halo star HD~88609.  Previous analyses of HD~88609 
\citep{jb02,j02,f00} show that this star is typical of very metal-poor
halo giants and has stellar parameters similar to those of D119.  In the 
re-analysis, we adopt primarily the equivalent width measurements of HD~88609 
from the three papers cited above.  Additional lines were measured from the
Keck/HIRES spectra kindly provided by Jennifer Johnson.

	The abundance analysis is conducted using the LTE code MOOG
\citep{s73} and Kurucz\footnote{http://kurucz.harvard.edu/} atmospheres
that do not include overshooting.  Using a model with overshooting raises
most of the $\log{\epsilon}$ values of all the neutral species by 
$\sim 0.1$ dex, but has very little effect on the ionized species.  

\citet{sbs} uses stellar parameters based on Fe lines, but \citet{ki03} 
show that while \teff{} values based on the excitation plot of \ion{Fe}{1} may
be reliable, surface gravities based on the ionization equilibrium
of Fe are not reliable for metal-poor globular cluster
giants.  For this star, we adopt a \teff{} value of 4440 K,
derived from the \citet{a99a} B-V calibration, assuming a reddening to
the Draco dSph of 0.03 mag \citep{m98}.  If a distance 
modulus of 19.58 to Draco is assumed \citep{m98}, then D119 has M$_V = -2.2$.
From Equation 1 of \citet{ki03} and the bolometric corrections of \citet{a99a}, 
the resulting \logg{} value is 0.88.  If the Shetrone et al. 
\teff{} value of 4370 K is used, the \logg{} value only drops to 0.86.
This is considerably higher than the Shetrone et al. \logg{} of 0.3.
When our value of \logg{} is used, the difference
log n(\ion{Fe}{1}) - log n(\ion{Fe}{2}) is $-0.02$ dex. 

We find a microturbulent velocity of $2.4 \pm 0.1$ km/s by 
forcing the abundance given by \ion{Fe}{1} lines to be independent of 
line strength.  The uncertainty in the mictroturbulent velocity, $v_t$,
was computed by multiplying the uncertainty in the slope of the least-squares
fit to the $\log{\epsilon{\rm(Fe\;I)}}$ vs. $\log{\rm (Reduced\;Width)}$ plot 
by the inverse of the change in this slope as a function of $v_t$.

For HD~88609, we adopt the \citet{a99b} 
\teff{} value of 4600 K.  \citet{h98} found M$_V = -1.2$ for this
star, which results in a \logg{} value of 1.38.  
We determine a microturbulent velocity of $2.1 \pm 0.1$ km/s.

Uncertainties for the abundances are calculated assuming the errors in 
stellar parameters and equivalent widths are independent.  The uncertainty
from parameter errors are approximated by changing the
\teff{} value by 150 K and altering the \logg{} and other parameters 
accordingly.  The uncertainty from line measurements is based on the 
calculations discussed above.  For D119, the dominant contribution to
error in the abundance determination comes from errors in the equivalent
widths, while for HD~88609, uncertainties in the stellar parameters
are the dominant contribution to the error in the abundance determination.

The final abundances, uncertainties, and upper limits for D119 and HD~88609 
are given in
Table 2.  As suggested by \cite{ki03}, we have used the
\ion{Fe}{1} abundance in determining the [X/Fe] ratio from neutral species
and the \ion{Fe}{2} abundance for singly-ionized species and [\ion{O}{1}].  
One of the advantages of using this method for computing abundance ratios
is that there is a reduced sensitivity to parameter changes.  

\section{D119 and the Early Enrichment}

Strontium and barium have the strongest absorption lines
for neutron-capture elements.  In HD~88609, the $\lambda4077$
\AA{} and $\lambda4215$ \AA{} \ion{Sr}{2} lines have equivalent widths of 167 m\AA{} and 160 m\AA{} and the $\lambda4554$
\ion{Ba}{2} line has an equivalent width of 91 m\AA{}.  We did not detect 
these two lines, nor any lines of any element heavier than Ni in D119.
Figure 1 shows spectral regions around the $\lambda4215$ \AA{} Sr II and 
$\lambda4554$ \AA{} Ba II lines for D119 and HD~88609.  

The difference in Sr and Ba abundances in these two stars is clear from the 
figure, and is confirmed by the abundance analysis.
The [Sr/Fe] and [Ba/Fe] abundance limits are about a factor 
of 100 less than what is seen in HD~88609, and the [Ba/Fe] ratio is below the
observed value for any metal-poor star.  The upper limits of the [Y/Fe] and 
[Zr/Fe] ratios are about a factor of 3 to 4 lower than similar metallicity 
galactic halo stars.  

The [Sr/Fe] and [Ba/Fe] ratios of D119, HD~88609, other Milky Way stars,
and the entire sample of literature satellite dwarf galaxy stars published
to date are plotted in Figure 2 as a function of [Fe/H].  The upper limits
of these ratios for D119 lie far outside the range of values for Milky Way
stars.  

D119 is not the only Draco member with a low [Ba/Fe] ratio.  The star
Draco 24,  at [Fe/H] $= -2.36$, lies about 0.5 dex below Milky Way 
stars of similar metallicity.  More metal-rich stars in Draco show ``normal''
[Ba/Fe] ratios.  While it is difficult to prove a trend with just two stars,
this may indicate that the whole of Draco started off as Ba-difficient, 
and whatever increased the metallicity of the star-forming gas was accompanied
by events that created Ba.  At high enough metallicity, the initial difficienty
became negligable.  It would be of interest to determine whether Draco 24 also
has a low [Sr/Fe] ratio.  

The Sextans star, S49 is of similar metallicity to D119 ([Fe/H] $= -2.85$),
but has a [Ba/Fe] value similar to that of Milky Way stars.
Whatever caused the lack of Ba (and by extention, Sr) within Draco did not
affect a similar star in the Sextans dSph galaxy.  

\subsection{Comparison to Type II Supernova Models}

The production of light elements in Type II supernovae  
is believed to be a strong function of the progenitor mass.  Heavier
progenitors should produce more oxygen and magnesium with respect to
calcium and titanium.  For example, the Type II models of \citet{n97},
assuming all the ejecta is well-mixed
before forming new stars, predict a [Mg/Ca] ratio of $+0.47$ for a 
70 \msun{} progenitor and a ratio of $-0.42$ for a 13 \msun{} progenitor.

The [Mg/Ca] ratio of D119 is $+0.61$, while it is $+0.26$ in HD~88609.
The [Mg/Fe], [Ca/Fe], and [Mg/Ca] ratios of these two stars, Milky Way
stars, and other dSph stars are shown as a function of [Fe/H] in Figure 3.
The most metal-poor ([Fe/H] $< -2$) dSph stars, including D119, show [Mg/Fe] 
ratios like Mikly Way stars, but nearly all the dSph stars show lower [Ca/Fe]
ratios.  However, nearly all the dSph stars show ``normal'' [Mg/Fe] ratios,
with D119 being noticably high.  It should be noted that all of the dSph
stars, including D119, were analyzed using a nearly identical line list and
have similar stellar parameters.  This reduces the effects of systematic
errors in the relative abundances.  The largest uncertainties in the 
relative abundances come from the line measurement uncertainty in the lower
S/N data. 

However, within the uncertainties of the measurement, the [Mg/Ca] ratio
of D119 is consistent with enrichment by only Type II supernovae with progenitor
masses of greater than $20-25$ \msun.  This would predict [O/Fe] and [Si/Fe] 
values for D119 of $\approx 0.5$.  Assuming a simple, well-mixed system, 
the [Mg/Ca] ratio in HD~88609 and of all of the other stars in Figure 3 
require the addition of ejecta from progenitors of initial lower mass.

In other words, all the stars have similar high-mass contributions
as evidenced by their high [Mg/Fe] ratios. 
The difference in [Mg/Ca] ratios between the stars therefore
depends on the prediction that lower-mass supernovae are significant
contributors of Ca.  Thus one would predict that D119 would show lower
abundances in other elements that are made in significant amounts by low-mass
progenitors.  

From the \citet{n97} models, these elements include Ti, and most of the
Fe-group elements (namely, Cr, Mn, Co, Ni, Cu, and Zn).  However, interpreting
these results is difficult because any combination of the \citet{n97} models 
cannot reproduce the observed abundances of metal-poor stars for several of
these elements.  For example, all progenitor masses produce sub-solar
[Ti/Fe] ratios although the vast majority of metal-poor stars show super-solar
[Ti/Fe] ratios.  This problem is a reminder of the speculative nature of using 
nucleosynthesis models to interpret abundance observations.  

\citet{un03} explore how progenitor mass, explosion energy and mixing/fallback
affect the [(Zn, Co, Mn, Cr)/Fe] ratios in ejecta from zero-metallicity
supernovae.  One result is that higher mass progenitors should have lower
[Zn/Fe] ratios, but mixing/fallback and high explosion energy enhance the
[Zn/Fe] ratio.  While these ratios in D119 may be systematically
lower than those seen in HD~88609, the uncertainties are large.  An improved
analyses with higher S/N spectra could utilize this diagnostic.

Alternatively, the pattern of high Na, Mg, and Al, but low Ca and Ti in D119 
is similar to what \citet{cl02} and \citet{i01} found for
supernova models with high C abundances left by central He burning.
However, it is difficult to understand how environment (e.g., in Draco vs. the
Milky Way) would affect the central burning properties of massive stars, so
this alone could not explain the abundance differences. 

A final difficulty in interpreting the nucleosynthesis models is 
the fate of the massive stars.  \citet{h03} find that for the
most metal-poor supernovae, those with masses of about 25 to 40 \msun{}
will leave a black hole formed from material falling back after the 
explosion.  
More massive stars (up to the mass range of hypernovae)
do not explode at all and collapse directly into black holes.  If significant
amounts of material are falling back onto the remnant for stars of even 
25 \msun{}, then the interpretation of the results becomes more dependent
on the explosion model.  

If it can be confirmed that D119 was solely enriched by the ejecta from 
high-mass supernovae ($M > 20$ to 25 \msun), then this would help eliminate
these stars as a site of the r-process.  \citet{m92}, \citet{t99} and 
\citet{w03} suggest that the r-process primarily takes place in lower mass 
(8--10 \msun) supernovae.  If these theories are correct, D119 may have
formed out of the ejecta of higher-mass supernovae before the
slower-evolving lower-mass progenitors could evolve.

These results imply the material in D119 came from stars at 
least as massive as those which \citet{t00} believe are the 
primary sites of the r-process.
The [Mg/H] ratio of D119 is $-2.41$, which is right at the point where
significant r-process contributions take place in their model.  
\citet{ts02} believe that the amount of hydrogen 
swept up by supernovae in dSph galaxies should be smaller, so a given supernova
shell would have higher metal-to-hydrogen ratios than in the Galactic halo.  
This would make D119 the product of lower-mass supernova.  However, 
metal-to-metal ratios, such as [Mg/Ca], should not be affected by the
amount of hydrogen swept up.

\subsection{Comparison to Prompt Enrichment Models}

The abundance pattern of metal-poor halo stars like HD~88609 are well-fit by 
the Qian \& Wasserburg models.  The contribution is dominated by the prompt 
and low-frequency events, with some high-frequency contributions to 
produce the Eu seen in these stars.  These components cannot reproduce 
the abundance pattern of D119.  In Figure 4(a) we plot the difference in 
$\log{\epsilon}$ between the Qian \& Wasserburg models.

All of the models overproduce Si, Ca, and Ti.  The low-frequency model is
included because it is predicted to produce no heavy neutron-capture elements, 
but the L-model produces no C and too much Cr and Mn.  The P-model also
produces too much Sr and possibly too much C, O, Y, Zr, and Zn.  There is
no combination of the two models that can be made to create a satisfactory
match to the abundance pattern of D119, mainly due to the overproduction of Sr
by the P-model.  Only the H-model produces any Ba, so the non-detection
of that element eliminates any sizable H-model contribution to D119.


Some of the differences seen in Figure 4(a) may be due to systematic differences
(potentially on the order a few tenths of a dex) between the abundance analysis 
applied here and the previous analyses used to calculate the Qian \& Wasserburg 
models.  A systematic analysis might show the abundance ratios of D119 are 
roughly consistent with the L-model predictions.

If some dSph galaxies did not undergo a 
prompt enrichment episode, then other very metal-poor stars in these galaxies
should show abundance patterns similar to that of D119.  
If these small 
galaxies are the CDM building blocks of larger galaxies, then these r-process 
poor stars should exist around the Milky Way.  Therefore, it is of great 
interest to 
determine if other dSph galaxies have r-process poor stars and whether there 
are accreted versions of D119 within the Galaxy.  

Zero-metallicity pair-instability hypernovae and normal supernovae have
been suggested as the source of any prompt enrichment.
Figure 4(b) is a comparison between the \citet{hw02} hypernovae models
and D119, plotting the scaled models for He core masses of 75, 100, and 
125 \msun.  One characteristic of the nucleosynthesis of pair production 
supernovae is the pronounced odd-even effect.  This pattern would make it
difficult for any combination of models to reproduce the abundances of D119.
For example, the models all overproduce Cr and underproduce Sc.  The 
zero-metallicity models of both \citet{un02} for both supernovae 
(13 to 30 \msun)
and hypernovae (150 to 270 \msun) also show this same strong odd-even effect.

\section{Another Low [Ba/Fe] Star:  BD~+80 245}

The unusual $\alpha$-poor nature of the metal-poor halo star BD~+80~245
was discovered by \citet{c97}.  Unlike most metal-poor stars
at [Fe/H] $\approx -2$, which have super-solar [$\alpha$/Fe] ratios (see
Figure 3), BD~+80 245 shows sub-solar ratios.  Further, \citet{f00} found 
that [Ba/Fe] and [Eu/Fe] are $\sim1$ and $\sim2$ dex below the ratios seen
in other halo stars of similar metallicity.

There are two possible explanations for the abundance pattern in BD~+80~245.
Since low-mass Type II supernovae produce lower $\alpha$/Fe ratios, the star
could be the product of a system only contaminated by low-mass events. 
However, the evidence provided here by D119 points to low-mass events
being producers of r-process elements, so BD~+80 245 should show high
[Ba/Fe] and [Eu/Fe] ratios, in contrast to observations.  The alternative
explanation, that BD~+80 245 is formed out of some Type II ejecta mixed
with Type Ia ejecta, is a more likely explanation for the abundance pattern 
found in this star.  \citet{i03} further explores the possible
enrichment history of this class of metal-poor, $\alpha$-poor stars.

\section{Summary}

We have analyzed new high S/N spectra of the metal-poor Draco dSph star D119,
confirming ([Fe/H]=$-2.96$) that this star remains
the most metal poor star presently known in any dwarf galaxy.
Our spectra of D119 show {\it no measurable lines for any neutron
capture elements}, a phenomenon not previously noted in any star.
One other star nearly as metal poor in the Sextans dwarf spheroidal
does not have the peculiar composition of D119.
While binary mass transfer processes can account for excesses of
neutron capture elements, we can imagine no astrophysical process
which might deplete so completely the heavy elements.  The unique
composition of D119 must be attributed to the chemical enrichment
history of the gas from which it was formed.

The abundance ratios of Draco 119 are consistent with that
expected for material formed from the ejecta of
massive Type II supernovae, a hypothesis that would also
be consistent with the great age and low metallicity of the Draco
dwarf spheroidal galaxy \citet{grill98}.
This is consistent with the predictions
of \citet{m92}, \citet{t99} and \citet{w03}, which places the
r-process in low mass Type II supernovae, and may conflict
with the prediction of \citet{t00} who propose intermediate
to massive Type II SNe as the site of the the r-process.  The
heavy element abundance distribution disagrees with the prompt enrichment
predictions of Qian \& Wasserburg and the pair-production supernovae
models of \citet{hw02} and \citet{un02}.  At present we find
no supernova model with predicted yields that can fit our observations.

\acknowledgements

The authors would like to thank J. Johnson and M. Shetrone for granting us
access to their spectra.  We would also like to thank the referees, S. Ryan
and M. Shetrone,
plus G. Wasserburg, K. Nomoto, C. Sneden, Wal Sargent and Brad Hansen for their 
many valuable suggestions.  We are especially grateful to the staff of Keck 
Observatory for their assistance, and S. Vogt and his team for the creation 
of HIRES.  JPF would like to thank Andy McWilliam for many useful
conversations.  RMR acknowledges support from grant AST-0098739 from the 
National Science Foundation.  The authors acknowledge the cultural role that
the summit of Mauna Kea has had within the indigenous Hawaiian community.
We are fortunate to have the opportunity to conduct observations from this
mountain.

\clearpage


\clearpage
\begin{deluxetable}{rcrrrrr}
\tablenum{1}
\tablewidth{0pt}
\tablecaption{Line List}
\tablehead{
\colhead{Wavelength} & \colhead{Ion} & \colhead{E.P.} & \colhead{log gf} &
\multicolumn{2}{c}{D119} & \colhead{HD 88609} \\
\colhead{\AA} & \colhead{} & \colhead{eV} & \colhead{} &
\multicolumn{2}{c}{EW m\AA} & \colhead{EW m\AA} \\
}
\startdata
4288.74 & CH\tablenotemark{a} & 0.64 &$-1.138$&   &  17 &  19 \\
4288.74 & CH    & 0.64 &$-1.115$&   &  bl &  bl \\
4307.31 & CH    & 0.16 &$-1.475$&   &  29 &  22 \\
4310.09 & CH    & 0.10 &$-1.534$&   &  28 &  40 \\
4310.11 & CH    & 0.10 &$-1.595$&   &  bl &  bl \\
4313.59 & CH    & 0.02 &$-1.923$&   &  30 &  21 \\
4313.65 & CH    & 0.02 &$-1.923$&   &  bl &  bl \\
6300.31 & [O I] & 0.00 &$-9.750$&$<$&   8 & \nodata \\
5889.97 &  Na I & 0.00 & hfs\tablenotemark{b} &   & 190 & 180 \\
4703.00 &  Mg I & 4.34 &$-0.520$&   &  56 &  61 \\
5172.70 &  Mg I & 2.71 &$-0.381$&   & 192 & 202 \\ 
5183.62 &  Mg I & 2.72 &$-0.170$&   & 219 & 214 \\
5528.42 &  Mg I & 4.34 &$-0.500$&   &  68 &  64 \\
3961.54 &  Al I & 0.00 & hfs    &   & 140 & 134 \\
4102.94 &  Si I & 1.91 &$-3.140$&$<$&  51 &  75 \\
5684.52 &  Si I & 4.93 &$-1.650$&$<$&  10 &   7 \\
5708.41 &  Si I & 4.95 &$-1.470$&$<$&  10 &   6 \\
4318.65 &  Ca I & 1.90 &$-0.210$&   &  21 &  44 \\
6122.23 &  Ca I & 1.89 &$-0.320$&   &  45 &  59 \\
6439.08 &  Ca I & 2.52 &$ 0.390$&   &  31 &  52 \\
4320.75 & Sc II & 0.61 &$-0.250$&   &  97 &  82 \\
4354.61 & Sc II & 0.61 &$-1.580$&   &  25 &  22 \\
4670.41 & Sc II & 1.36 &$-0.580$&   &  25 &  24 \\
5239.82 & Sc II & 1.45 &$-0.760$&   &  22 &  13 \\
5526.82 & Sc II & 1.77 &$+0.020$&   &  46 &  30 \\
4533.24 &  Ti I & 0.85 &$+0.540$&   &  27 &  43 \\
4534.78 &  Ti I & 0.84 &$+0.340$&   &  34 &  34 \\
4555.49 &  Ti I & 0.85 &$-0.430$&   &  15 &   9 \\
4981.73 &  Ti I & 0.85 &$+0.560$&   &  49 &  22 \\
4999.50 &  Ti I & 0.83 &$+0.310$&   &  29 &  39 \\
5014.24 &  Ti I & 0.81 &$+0.110$&   &  48 &  \nodata \\
5020.03 &  Ti I & 0.84 &$-0.350$&   &  10 &  12 \\
5022.87 &  Ti I & 0.83 &$-0.370$&   &  19 &  \nodata \\
5039.96 &  Ti I & 0.21 &$-1.070$&   &  29 &  24 \\
5064.65 &  Ti I & 0.05 &$-0.930$&   &  41 &  30 \\
5173.74 &  Ti I & 0.00 &$-1.060$&   &  22 &  26 \\
5192.97 &  Ti I & 0.02 &$-0.950$&   &  34 &  32 \\
5210.39 &  Ti I & 0.05 &$-0.820$&   &  24 &  34 \\
4394.06 & Ti II & 1.22 &$-1.770$&   &  52 &  54 \\
4395.04 & Ti II & 1.08 &$-0.510$&   & 121 & 137 \\
4395.84 & Ti II & 1.24 &$-1.970$&   &  39 &  41 \\
4398.29 & Ti II & 1.22 &$-2.780$&   &  13 &  11 \\
4443.80 & Ti II & 1.08 &$-0.700$&   & 110 & 116 \\
4444.56 & Ti II & 1.12 &$-2.210$&   &  32 &  38 \\
4450.48 & Ti II & 1.08 &$-1.510$&   &  68 &  81 \\
4493.52 & Ti II & 1.08 &$-2.830$&   &  11 &  14 \\
4501.27 & Ti II & 1.12 &$-0.760$&   & 117 & 116 \\
4708.66 & Ti II & 1.24 &$-2.370$&   &  14 &  26 \\
4798.53 & Ti II & 1.08 &$-2.670$&   &  24 &  26 \\
5336.79 & Ti II & 1.58 &$-1.630$&   &  42 &  43 \\
5129.16 & Ti II & 1.89 &$-1.390$&   &  27 & \nodata \\
5154.07 & Ti II & 1.57 &$-1.920$&   &  16 & \nodata \\
5226.55 & Ti II & 1.57 &$-1.300$&   &  70 & \nodata \\
5381.01 & Ti II & 1.57 &$-2.080$&   &  23 & \nodata \\
4379.23 &   V I & 0.30 &$+0.550$&   &  44 &  20 \\
4389.98 &   V I & 0.28 &$+0.270$&   &  18 &  11 \\
4254.35 &  Cr I & 0.00 &$-0.110$&   & 107 & 109 \\
4616.13 &  Cr I & 0.98 &$-1.200$&   &  12 &  13 \\
4646.17 &  Cr I & 1.03 &$-0.720$&   &  20 &  26 \\
5296.70 &  Cr I & 0.98 &$-1.400$&   &  15 &  12 \\
5345.81 &  Cr I & 1.00 &$-0.980$&   &  19 &  21 \\
5409.80 &  Cr I & 1.03 &$-0.720$&   &  31 &  33 \\
4652.17 &  Cr I & 1.00 &$-1.030$&   &  23 &  17 \\
5206.04 &  Cr I & 0.94 &$+0.019$&   &  68 &  \nodata \\
5409.80 &  Cr I & 1.03 &$-0.720$&   &  36 &  33 \\
4030.76 &  Mn I & 0.00 & hfs    &   &  96 & 117 \\
4033.07 &  Mn I & 0.00 & hfs    &   &  89 & 109 \\
4034.49 &  Mn I & 0.00 & hfs    &   & 106 &  97 \\
4250.13 &  Fe I & 2.47 &$-0.370$&   &  68 &  84 \\
4260.47 &  Fe I & 2.40 &$+0.140$&   &  92 & 108 \\
4282.41 &  Fe I & 2.17 &$-0.780$&   &  83 &  82 \\
4337.05 &  Fe I & 1.56 &$-1.660$&   &  84 &  89 \\
4375.93 &  Fe I & 0.00 &$-2.990$&   & 126 & 126  \\
4404.75 &  Fe I & 1.56 &$-0.100$&   & 150 & 149 \\
4415.13 &  Fe I & 1.61 &$-0.620$&   & 128 & 128 \\
4427.31 &  Fe I & 0.05 &$-3.000$&   & 140 & \nodata \\
4430.62 &  Fe I & 2.22 &$-1.620$&   &  48 & \nodata \\
4443.20 &  Fe I & 2.86 &$-1.040$&   &  33 &  27 \\
4445.48 &  Fe I & 0.09 &$-5.400$&   &  12 &  10 \\
4447.73 &  Fe I & 2.22 &$-1.340$&   &  66 &  58 \\
4461.65 &  Fe I & 0.09 &$-3.170$&   & 127 & 122 \\
4489.74 &  Fe I & 0.12 &$-3.930$&   &  86 &  77 \\
4494.57 &  Fe I & 2.20 &$-1.100$&   &  53 &  76 \\
4531.15 &  Fe I & 1.49 &$-2.110$&   &  80 &  75 \\
4602.94 &  Fe I & 1.49 &$-2.180$&   &  83 &  14 \\
4630.13 &  Fe I & 2.28 &$-2.590$&   &  19 &   9 \\
4632.92 &  Fe I & 1.61 &$-2.900$&   &  41 &  24 \\
4647.43 &  Fe I & 2.94 &$-1.350$&   &  31 &  16 \\
4733.60 &  Fe I & 1.49 &$-2.950$&   &  37 &  31 \\
4736.77 &  Fe I & 3.20 &$-0.750$&   &  37 &  30 \\
4859.74 &  Fe I & 2.86 &$-0.760$&   &  59 & \nodata \\
4871.32 &  Fe I & 2.85 &$-0.360$&   &  79 &  64 \\
4872.14 &  Fe I & 2.87 &$-0.570$&   &  72 &  52 \\
4890.75 &  Fe I & 2.86 &$-0.390$&   &  74 &  62 \\
4891.49 &  Fe I & 2.84 &$-0.110$&   &  76 &  76 \\
4903.31 &  Fe I & 2.87 &$-0.930$&   &  39 &  38 \\
4918.99 &  Fe I & 2.85 &$-0.340$&   &  84 &  65 \\
4920.50 &  Fe I & 2.82 &$+0.070$&   &  83 &  89 \\
4924.77 &  Fe I & 2.28 &$-2.250$&   &  24 &  19 \\
4938.81 &  Fe I & 2.86 &$-1.080$&   &  32 &  28 \\
4939.69 &  Fe I & 0.86 &$-3.300$&   &  72 &  58 \\
4994.13 &  Fe I & 0.92 &$-3.040$&   &  76 &  70 \\
5006.12 &  Fe I & 2.83 &$-0.660$&   &  70 &  59 \\
5041.07 &  Fe I & 0.95 &$-3.090$&   &  78 &  71 \\
5049.82 &  Fe I & 2.28 &$-1.340$&   &  68 &  57 \\
5051.64 &  Fe I & 0.92 &$-2.760$&   &  92 &  90 \\
5068.77 &  Fe I & 2.93 &$-1.040$&   &  29 &  26 \\
5079.23 &  Fe I & 2.20 &$-2.030$&   &  31 &  34 \\
5079.74 &  Fe I & 0.99 &$-3.180$&   &  71 &  58 \\
5083.34 &  Fe I & 0.96 &$-2.920$&   &  86 &  77 \\
5123.72 &  Fe I & 1.01 &$-3.030$&   &  77 &  65 \\
5127.36 &  Fe I & 0.92 &$-3.270$&   &  69 &  61 \\
5141.73 &  Fe I & 2.42 &$-1.960$&   &  14 &  16 \\
5142.93 &  Fe I & 0.96 &$-3.080$&   &  65 &  69 \\
5151.92 &  Fe I & 1.01 &$-3.280$&   &  63 &  47 \\
5166.29 &  Fe I & 0.00 &$-4.160$&   & 100 &  84 \\
5171.60 &  Fe I & 1.49 &$-1.750$&   & 109 &  97 \\
5191.45 &  Fe I & 3.03 &$-0.550$&   &  52 &  44 \\
5192.34 &  Fe I & 2.99 &$-0.420$&   &  54 &  54 \\
5194.94 &  Fe I & 1.56 &$-2.050$&   &  87 &  79 \\
5198.71 &  Fe I & 2.22 &$-2.090$&   &  24 &  25 \\
5216.28 &  Fe I & 1.61 &$-2.110$&   &  78 &  71 \\
5217.39 &  Fe I & 3.21 &$-1.070$&   &  25 &  14 \\
5225.53 &  Fe I & 0.11 &$-4.750$&   &  55 &  35 \\
5232.94 &  Fe I & 2.94 &$-0.100$&   &  83 &  77 \\
5247.05 &  Fe I & 0.09 &$-4.910$&   &  38 &  26 \\
5250.21 &  Fe I & 0.12 &$-4.900$&   &  40 &  28 \\
5269.54 &  Fe I & 0.86 &$-1.330$&   & 173 & 161 \\
5281.79 &  Fe I & 3.03 &$-0.830$&   &  34 &  30 \\
5283.62 &  Fe I & 3.23 &$-0.520$&   &  47 &  39 \\
5302.30 &  Fe I & 3.28 &$-0.720$&   &  28 &  26 \\
5307.37 &  Fe I & 1.61 &$-2.950$&   &  34 &  24 \\
5324.19 &  Fe I & 3.21 &$-0.100$&   &  55 & \nodata \\
5332.90 &  Fe I & 1.55 &$-2.780$&   &  50 &  34 \\
5339.93 &  Fe I & 3.27 &$-0.650$&   &  27 &  27 \\
5341.02 &  Fe I & 1.60 &$-1.950$&   &  98 &  83 \\
5371.50 &  Fe I & 0.96 &$-1.644$&   & 157 & 141 \\
5383.37 &  Fe I & 4.31 &$+0.640$&   &  33 &  24 \\
5393.17 &  Fe I & 3.24 &$-0.710$&   &  28 &  25  \\
5397.13 &  Fe I & 0.92 &$-1.950$&   & 158 & 135 \\
5404.15 &  Fe I & 4.42 &$+0.520$&   &  20 &  27 \\
5405.78 &  Fe I & 0.99 &$-1.800$&   & 144 & 132 \\
5415.20 &  Fe I & 4.37 &$+0.640$&   &  26 &  20 \\
5429.70 &  Fe I & 0.95 &$-1.880$&   & 150 & 134 \\
5434.53 &  Fe I & 1.01 &$-2.080$&   & 126 & \nodata \\
5501.46 &  Fe I & 0.95 &$-3.050$&   &  85 &  77 \\
5506.78 &  Fe I & 0.99 &$-2.700$&   &  99 &  89 \\
5615.66 &  Fe I & 3.33 &$+0.050$&   &  71 & \nodata \\
6136.62 &  Fe I & 2.45 &$-1.400$&   &  56 &  52 \\
6137.70 &  Fe I & 2.59 &$-1.366$&   &  59 &  43 \\
6191.57 &  Fe I & 2.43 &$-1.416$&   &  70 & \nodata \\
6219.29 &  Fe I & 2.20 &$-2.438$&   &  37 &  19 \\
6230.74 &  Fe I & 2.56 &$-1.276$&   &  73 & \nodata \\
6252.57 &  Fe I & 2.40 &$-1.757$&   &  57 &  42 \\
6421.36 &  Fe I & 2.28 &$-2.014$&   &  30 &  35 \\
6430.86 &  Fe I & 2.18 &$-1.946$&   &  45 &  45 \\
4508.29 & Fe II & 2.86 &$-2.330$&   &  40 &  42 \\
4515.34 & Fe II & 2.84 &$-2.480$&   &  24 &  31 \\
4555.89 & Fe II & 2.83 &$-2.390$&   &  40 &  39 \\
5018.45 & Fe II & 2.89 &$-1.220$&   & 101 & 106 \\
5197.56 & Fe II & 3.23 &$-2.100$&   &  35 &  27 \\
5234.62 & Fe II & 3.22 &$-2.230$&   &  27 &  36 \\
5276.00 & Fe II & 3.20 &$-1.940$&   &  41 &  38 \\
4121.32 &  Co I & 0.92 & hfs    &   &  82 &  77 \\
5035.36 &  Ni I & 3.63 &$+0.290$&   &  14 &  12 \\
5137.08 &  Ni I & 1.68 &$-1.990$&   &  13 &  20 \\
5476.91 &  Ni I & 1.83 &$-0.890$&   &  65 &  64 \\
5105.54 &  Cu I & 1.39 &$-1.720$&$<$&   9 & \nodata \\
4810.55 &  Zn I & 4.08 &$-0.170$&$<$&  10 &  15 \\
4077.71 & Sr II & 0.00 & hfs    &$<$&  53 & 169 \\
4215.52 & Sr II & 0.00 & hfs    &$<$&  25 & 160 \\
4883.69 &  Y II & 1.08 &$+0.070$&$<$&  10 &  20 \\
4900.11 &  Y II & 1.03 &$-0.090$&$<$&  10 &  21 \\
5087.42 &  Y II & 1.08 &$-0.170$&$<$&   9 &  12 \\
4208.98 & Zr II & 0.71 &$-0.460$&$<$&  25 &  30 \\
4554.03 & Ba II & 0.00 & hfs    &$<$&  14 &  91 \\
4934.08 & Ba II & 0.00 & hfs    &$<$&  10 & \nodata \\
4061.09 & Nd II & 0.47 &$+0.300$&$<$&  35 &   5 \\
4109.46 & Nd II & 0.32 &$+0.180$&$<$&  35 &   5 \\
4129.72 & Eu II & 0.00 & hfs    &$<$&  41 &   9 \\
\enddata
\tablenotetext{a}{CH line analysis assumed a disassociation energy of
3.47~eV.  Lines with ``bl'' in the EW columns were lines analyzed as
blends with the line preceeding it in the list, with the given EW
assumed to cover both lines.}
\tablenotetext{b}{Lines with log gf values marked as ``hfs'' were treated 
as blend of many hyperfine components.   
Hyperfine data taken from Johnson (2002), McWilliam (1998), and
McWilliam et al. (1995).}
\end{deluxetable}


\clearpage
\begin{deluxetable}{crrrrrr}
\tablenum{2}
\tablewidth{0pt}
\tablecaption{Observed Abundances}
\tablehead{
\colhead{} & \multicolumn{3}{c}{Draco 119} & \multicolumn{3}{c}{HD 88609}\\
\colhead{Species} & \multicolumn{1}{r}{log $\epsilon$(X)} & \colhead{[X/Fe]\tablenotemark{a}} & \colhead{$\sigma$} &\multicolumn{1}{r}{log $\epsilon$(X)} &  \colhead{[X/Fe]\tablenotemark{a}} & \colhead{$\sigma$}
}
\startdata
C\tablenotemark{b} & $5.18$ & $-0.48$ & $0.26$ & $5.19$ & $-0.43$ & $0.14$ \\
{\rm [O I]} & $< 6.6$ & $< 0.8$  & \nodata & \nodata & \nodata & \nodata \\
Na I  & $3.74$  & $+0.36$  & $0.12$  & $3.53$ & $+0.14$ & $0.07$ \\
Mg I  & $5.17$  & $+0.54$  & $0.22$  & $5.23$ & $+0.59$ & $0.07$ \\
Al I  & $3.28$  & $-0.24$ & $0.49$   & $3.17$ & $-0.36$ & $0.10$ \\
Si I  & $< 4.8$ & $< +0.2$ & \nodata & $5.18$ & $+0.57$ & $0.11$ \\
Ca I  & $3.34$  & $-0.07$  & $0.17$  & $3.75$ & $+0.33$ & $0.08$ \\
Sc II & $0.43$  & $+0.30$  & $0.18$  & $0.40$ & $+0.13$ & $0.07$ \\
Ti I  & $2.11$  & $+0.07$  & $0.13$  & $2.22$ & $+0.17$ & $0.05$ \\
Ti II & $2.23$  & $+0.21$  & $0.15$  & $2.48$ & $+0.32$ & $0.06$ \\
V I   & $1.19$  & $+0.14$  & $0.29$  & $0.94$ & $-0.12$ & $0.05$ \\
Cr I  & $2.30$  & $-0.42$  & $0.14$  & $2.45$ & $-0.28$ & $0.05$ \\
Mn I  & $1.75$  & $-0.69$  & $0.48$  & $2.15$ & $-0.30$ & $0.11$ \\
Fe I  & $4.57$  & $-2.95$  & $0.20$  & $4.58$ & $-2.94$ & $0.14$ \\
Fe II & $4.55$  & $-2.97$  & $0.12$  & $4.69$ & $-2.83$ & $0.05$ \\
Co I  & $< 2.1$ & $< +0.1$ & \nodata & $2.10$ & $+0.12$ & $0.07$ \\
Ni I  & $3.15$  & $-0.15$  & $0.17$  & $3.31$ & $+0.00$ & $0.06$ \\
Cu I  & $< 1.17$  & $< -0.1$ & \nodata & \nodata & \nodata & \nodata \\
Zn I  & $< 1.7$ & $< +0.1$ & \nodata & $2.01$ & $+0.35$ & $0.13$ \\
Sr II & $<-2.6$ & $< -2.5$ & \nodata & $0.00$ & $-0.07$ & $0.12$ \\
Y II  & $<-1.3$ & $< -0.5$ & \nodata & $-0.80$ & $-0.21$ & $0.07$ \\
Zr II & $<-0.2$ & $< +0.1$ & \nodata & $0.04$ & $+0.27$ & $0.08$ \\
Ba II & $<-3.3$ & $< -2.6$ & \nodata & $-1.61$ & $-0.91$ & $0.10$ \\
Nd II & $<-0.7$ & $< +0.8$ & \nodata & $-1.62$ & $-0.27$ & $0.14$ \\
Eu II & $<-2.0$ & $< +0.4$ & \nodata & $-2.59$ & $-0.50$ & $0.08$ \\
\enddata
\tablenotetext{a}{[Fe/H] ratios given for Fe I and Fe II.}
\tablenotetext{b}{Based on G-band CH lines.}
\end{deluxetable}

\clearpage
\figcaption{Two sample spectral regions showing the difference in 
neutron-capture line strengths between D119 (solid) and 
the halo star HD~88609 (dotted), which has the same metallicity as D119.
The left
panel shows the Sr II 4215~\AA{} line for both stars.  Note the near identical
strength of the Fe I line.   The right panel shows the Ba II 4554~\AA{} line
for both stars.\label{fig1}}  

\figcaption{The [Sr/Fe] and [Ba/Fe] ratios of D119, HD~88609, field halo
stars, and dSph giants plotted against [Fe/H].  The field star data are from
McWilliam (1995), Ryan et al. (1996), Burris et al. (2000), Fulbright (2000), 
Johnson (2002) and Ivans et al. (2003).  All of the field stars (with the 
exception of BD~+80+245) are plotted using the same symbol to prevent 
confusion.  Except for D119, the dSph sample comes from Shetrone et al. (2001)
and Shetrone et al. (2003).  In both plots, the upper limits for D119 clearly
lie below other stars of similar metallicity.  \label{f2}} 

\figcaption{The [Mg/Fe], [Ca/Fe] and [Mg/Ca] ratios of D119 and the other
stars plotted in Figure 2 (same symbols used).  In the top panel, D119 and
the other very metal-poor dSph stars show [Mg/Fe] ratios similar to that
seen in the Milky Way population.  The dSph stars, however, show lower [Ca/Fe]
ratios in the middle panel.  In the bottom panel, D119 show a very high [Mg/Ca]
ratio, which is the result of having a slighly higher [Mg/Fe] ratio and 
slightly lower [Ca/Fe] ratio than similar stars (such as Sextans S49).
Overall, the dSph stars show [Mg/Ca] ratios not too different than the 
Milky Way sample, suggesting that with the exception of D119, the mass function 
of Type II supernovae progenitors that polluted the two groups was similar.
\label{f3}}

\figcaption{Comparison of theoretical models to the observed abundances of
D119.  In both panels, the quantity $\Delta{}\log{}\epsilon{}$ is defined 
as $\log{}\epsilon{\rm (Model)} - \log{}\epsilon{}{\rm (D119)}$, and
the models are scaled to the Fe abundance of D119.  Positive values 
indicate overproduction by the models, while the zero-line indicates a
perfect match to D119.  The upward-pointing arrows denote 
elements for which the marked points are lower limits based on the adopted 
upper limit of the abundance measured in D119.  
(a) Comparison of the P- and L-model predictions of Qian \& 
Wasserburg to the observed abundances of D119.
The prediction of the carbon abundance from the P-model is
an upper limit.
(b) Comparison of the Heger \& Woosley (2002) predictions of the
nucleosynthesis of pair instability supernovae to the observed abundances
of D119.  The three models are based on He-core masses of 75, 100 
and 125 \msun.\label{fig4}}

\begin{figure}
\plotone{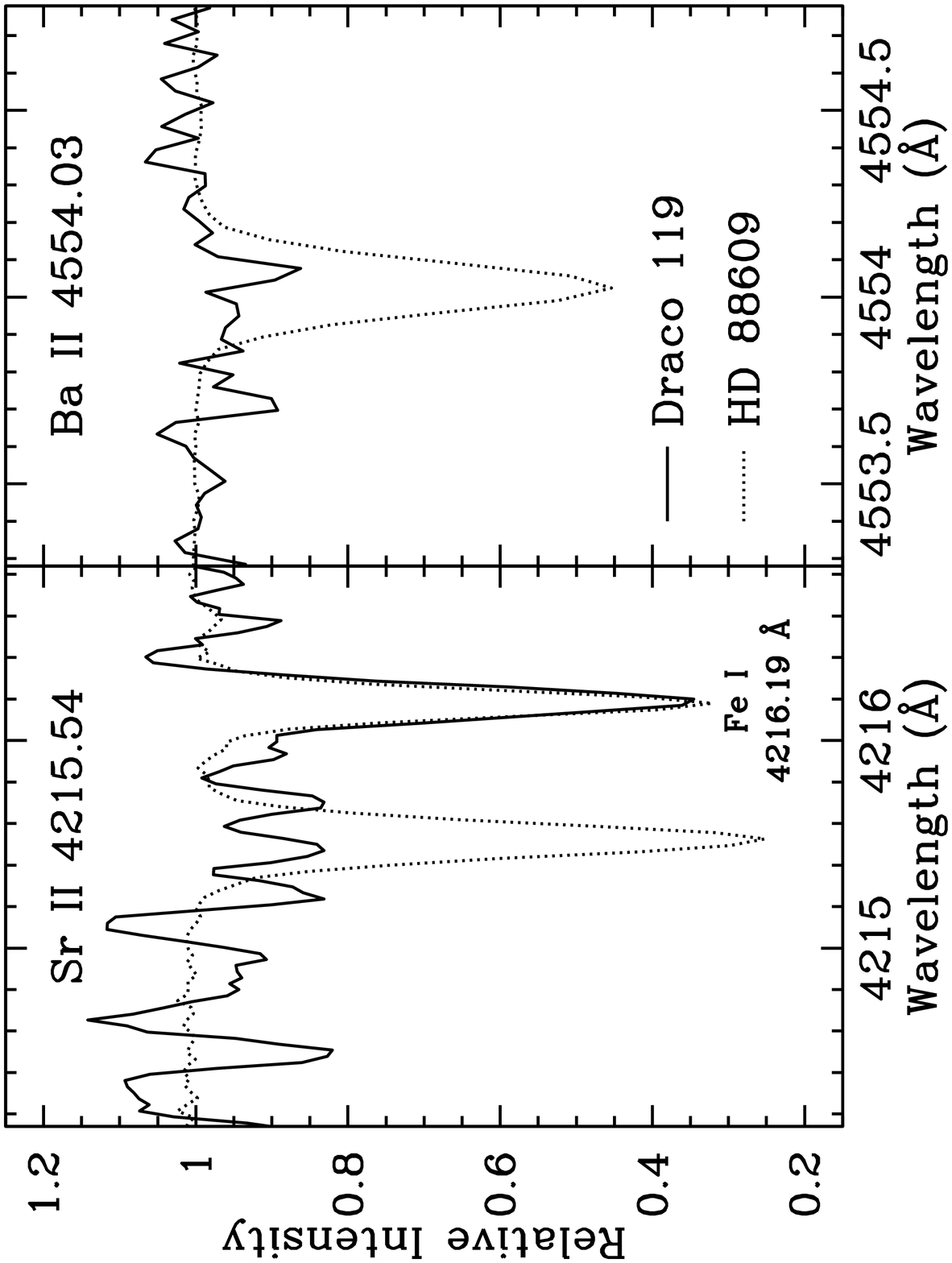}
\end{figure}

\clearpage

\begin{figure}
\plotone{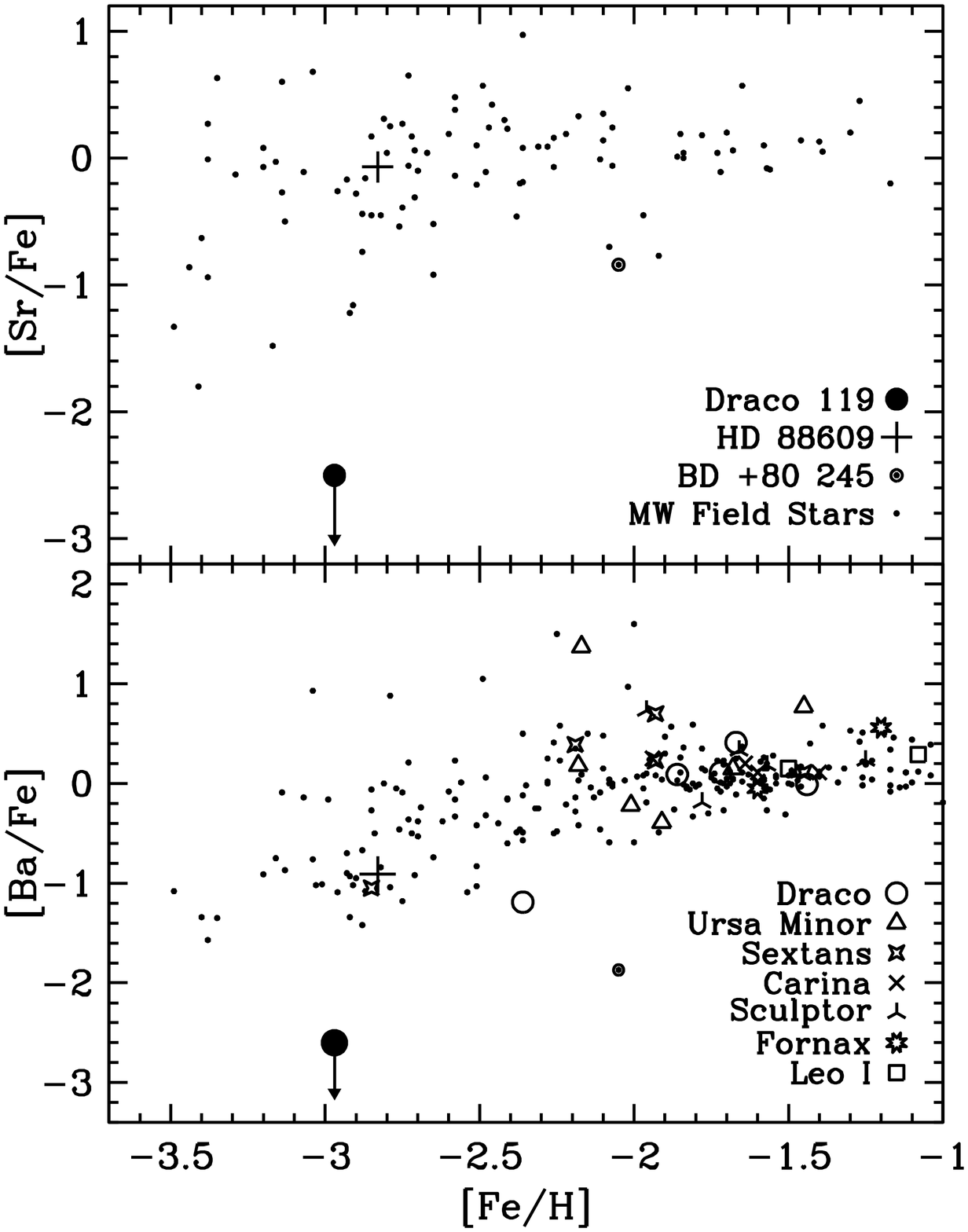}
\end{figure}

\clearpage

\begin{figure}
\plotone{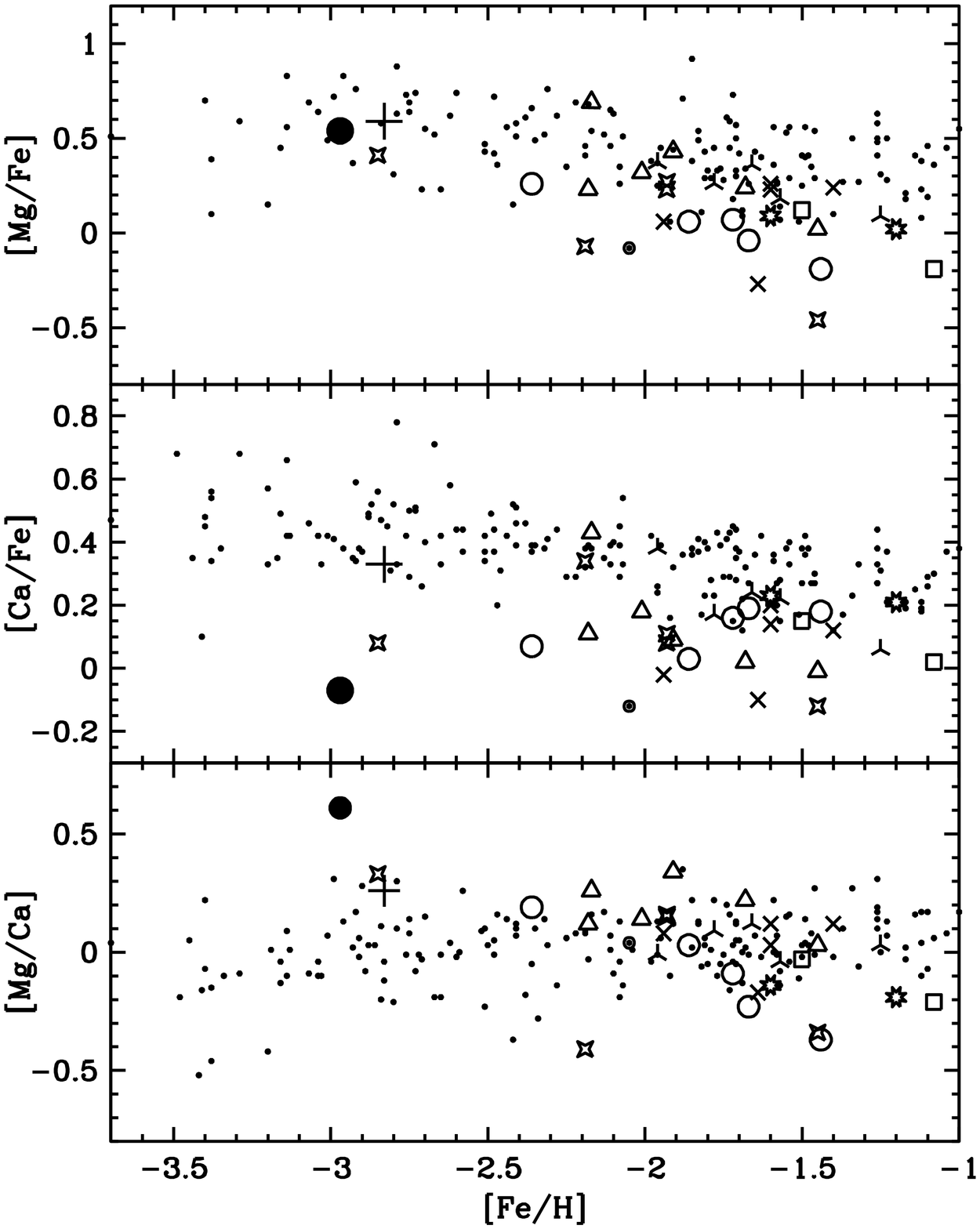}
\end{figure}

\clearpage

\begin{figure}
\plotone{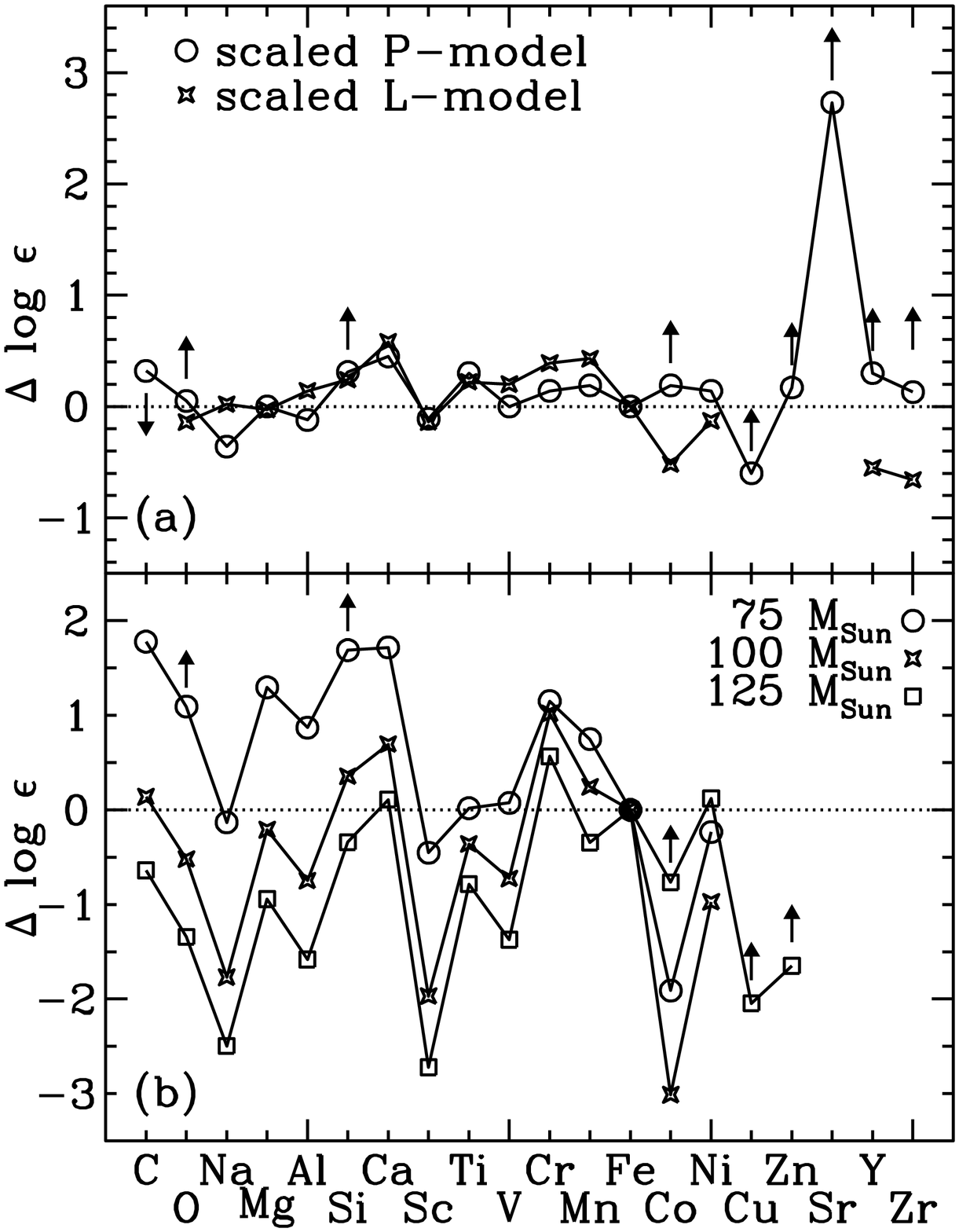}
\end{figure}

\end{document}